\documentclass[aps,prl,reprint,showpacs,floatfix,superscriptaddress]{revtex4-1}

\usepackage{graphicx}
\usepackage{amsmath,amsthm,amsfonts,amssymb,amscd}
\usepackage[latin9]{inputenc}
\usepackage{mathrsfs}
\usepackage{mathtools}
\usepackage{amsmath}
\usepackage{stmaryrd}

\usepackage{setspace}
\usepackage[T1]{fontenc}
\usepackage[american]{babel}

\begin{document}
\title{Quantum sensing of open systems: Estimation of damping constants and temperature}
\author{J. Wang}
\affiliation{Department of Physics and Astronomy, Texas A\&M University, College Station, Texas 77843, USA}
\author{L. Davidovich}
\affiliation{Department of Physics and Astronomy, Texas A\&M University, College Station, Texas 77843, USA}
\affiliation{Instituto de F\'\i sica, Universidade Federal do Rio de Janeiro, Rio de Janeiro, RJ 21941-972, Brazil}
\affiliation{Hagler Institute for Advanced Study and Institute for Quantum Science and Engineering, Texas A\&M University, College Station, Texas 77843, USA}
\author{G. S. Agarwal}
\affiliation{Department of Physics and Astronomy, Texas A\&M University, College Station, Texas 77843, USA}
\affiliation{Institute for Quantum Science and Engineering and Department of Biological and Agricultural Engineering, Texas A\&M University, College Station, Texas 77843, USA}

\begin{abstract}
We determine quantum precision limits for estimation of damping constants and temperature of lossy bosonic channels. A direct application would be the use of light for estimation of the absorption and the temperature of a transparent slab. Analytic lower bounds are obtained for the uncertainty in the estimation, through a purification procedure that replaces the master equation description by a unitary evolution involving the system and ad hoc environments. For zero temperature, Fock states are shown to lead to the minimal uncertainty in the estimation of damping, with boson-counting being the best measurement procedure. In both damping and temperature estimates, sequential pre-thermalization measurements, through a stream of  single bosons, may lead to huge gain in precision.  
\end{abstract}

\maketitle

{\it Introduction.} The quest for better precision in the estimation of parameters is common to many areas of science, ranging  from probing weak electric and magnetic fields, temperature, pressure, and small rotations and displacements, to high-resolution spectroscopy and magnetic resonance, with applications to atomic clocks, geophysics, medicine, and biology. Fundamental limits of precision have been established, within the realm of classical physics, by Cram\'er, Rao, and Fisher \cite{cramer,fisher}.  The usual procedure  involves measuring a probe, prepared in a convenient initial state, after it has interacted with the system under investigation, and then obtaining from the measurement results an estimation of the parameter of interest, through some convenient estimator. Through a generalization of the classical framework to quantum mechanics \cite{helstrom,holevo,caves,caves2}, it has been realized that quantum probes, prepared in states with features like squeezing and entanglement, help to increase the precision of the estimation, for the same amount of resources (which could be the number of atoms or photons used in the estimation). This has been relevant, for instance, for extending the coverage of gravitational-wave interferometers, with the use of squeezed light \cite{caves,gravitation} or of entangled states \cite{schnabel}, for increasing the magnetic sensitivity with spin squeezing \cite{sewell},  for optimal thermometry \cite{thermo}, for detecting weak electric fields with superpositions of Rydberg states \cite{facon}, for achieving quantum-enhanced contrast and resolution in biological microscopy \cite{bowen,taylor}, and for superresolution of spatial separation and frequency \cite{gefen}. Quantum sensing \cite{sensing,pezze} involves the exploration of subtle quantum effects to increase the precision of parameter estimation. Quantum sensors have become one of the most promising applications of quantum technologies \cite{diamond,nano,bouton}, involving single- or multi-parameter estimation \cite{multi1,multi2}.

The unavoidable interaction between these systems and their environments may reduce the advantage of using quantum states, due to the fragility of these resources in the presence of noisy processes, like damping and diffusion.  However, sometimes these processes may yield important information on the system. The damping rate of a particle moving in a medium may  allow the estimation of the quantum memory time and radiation properties \cite{henning}. Absorption spectroscopy has a wide range of applications, in remote sensing \cite{jenson}, in chemistry and atomic physics \cite{haswell}, in astronomy \cite{wit}, and in the characterization of materials, not only at the macroscopic level, but also for microscopic systems, like cells and organelles \cite{biology}. Moreover, tasks like the precise estimation of phases in an interferometer must necessarily include a precise estimation of photon damping and phase diffusion.

Here we derive the uncertainties in the estimation of both damping and temperature of a lossy bosonic channel, with boson-counting as the measurement procedure.  This is of great interest for several areas of science,  the most prominent example being the use of light to investigate absorption and temperature of samples \cite{thermometry,NIST}.  The precision in the estimation is limited both by the uncertainty in the number of bosons in the probe and by the noise introduced in the boson distribution by the probed system. This suggests that one should minimize the variance of the boson-number distribution of the probe, so incoming Fock states should render better results, as opposed to what happens in noiseless phase estimation, when the variance should be  maximized, for a given amount of resources (in this case, input photons). 

We discuss the advantages of using single-boson states and boson-counting measurements for damping and temperature estimation and compare our results within literature \cite{paris,braun,adesso}. Sequential pre-thermalization measurements with  single-boson streams are shown to lead to a huge increase in the precision. We also obtain analytic lower bounds for the uncertainty in the estimation of both damping and temperature, through a purification procedure that replaces the master equation description by a unitary evolution involving the system and ad hoc environments. These bounds are shown to be tight in two limiting cases, both involving boson-counting measurements: zero temperature for damping estimation, and vacuum input for temperature estimation. For other situations, and for the range of parameters here considered, they are very close to the exact numerical solutions. 

The usual procedure in parameter estimation consists in obtaining the uncertainty in the parameter, for a given initial state, from the Fisher information \cite{cramer,fisher}. For a complete set of measurement results $\{j\}$, on a probe that carries information about the parameter $X$ to be estimated, and for unbiased estimators, so that $\langle X\rangle$ equals the true value of the parameter, the standard deviation in the estimation of $X$ is given by the Cram\'er-Rao expression $\delta X\ge1/\sqrt{NF(\langle X\rangle)}$, where $F(X)$ is the Fisher information, given by $F(X)=\sum_j[1/P_j(X)][dP_j(X)/dX]^2$, $N$ is the number of repetitions of the experiment, and $P_j(X)$ is the probability of getting the experimental result $j$ if the value of the parameter is $X$. As shown by Fisher, the lower bound can be reached asymptotically for $N\rightarrow\infty$. The ultimate precision in the estimation of a parameter, for a given initial state, is obtained by maximizing $F(X)$ over all possible measurements: this defines the quantum Fisher information (QFI) ${\cal F}_Q(X)$. In the absence of noise, analytic expressions can be obtained for the QFI \cite{helstrom, holevo}.  For a parameter-dependent unitary evolution $U(X)$ of the probe, ${\cal F}_Q(X)$ is equal to four times the variance $(\Delta G)^2$, calculated in the initial state of the probe, with  $G\equiv i(dU^\dagger(X)/dX)U(X)$ being the generator of $U(X)$. However, this is not so for open systems, which require, in general, the diagonalization of the parameter-dependent density matrix of the probe, usually a cumbersome task for high-dimensional systems.
\par
A general method for obtaining an upper bound for the quantum Fisher information of an open system was introduced in \cite{escher}. It consists in purifying the open system, by considering the joint unitary evolution of system+environment. There is an infinite number of purifications, which must satisfy  the criterion that the reduced description of the system  -- obtained by tracing out the environment -- should coincide with the one given by the  master equation. The quantum Fisher information of the purified system should be larger or at least equal to the QFI of the system, since allowing measurements on system+environment yields no less information on the parameter than measuring the system alone. If the environment is chosen in such a way that measurements on system+environment do not give more information than measurements on the system, the corresponding upper bound is tight. In \cite{escher}, it was shown that this can always be accomplished. Finding the best purification could provide therefore an alternative to the involved procedures that deal directly with the open system. This method has led to exact solutions for the estimation of forces acting on damped harmonic oscillators \cite{camille} and very good approximations for the estimation of transition frequencies in atomic spectroscopy in the presence of dephasing, and phases in optical interferometers, subject to damping \cite{escher} and diffusion \cite{variational}. In the following, this method is applied to the estimation of damping and temperature with bosonic probes.

\noindent {\it Estimation of damping}. Boson damping can be described by the master equation
\begin{align}\label{master}
{d\rho\over dt}=\gamma(n_T+1)(2a\rho a^\dagger-a^\dagger a\rho-\rho a^\dagger a)\nonumber\\
+\gamma n_T(2a^\dagger\rho a -a a^\dagger\rho-\rho aa^\dagger )\,,
\end{align}
where $\rho$ is the density matrix of the bosonic probe, $\gamma$ is the damping constant, $n_T$ is the number of thermal bosons, and $a$ and $a^\dagger$ are boson annihilation and creation operators, with $[a,a^\dagger]=1$.

\begin{figure}[tb]
\includegraphics[width=\columnwidth]{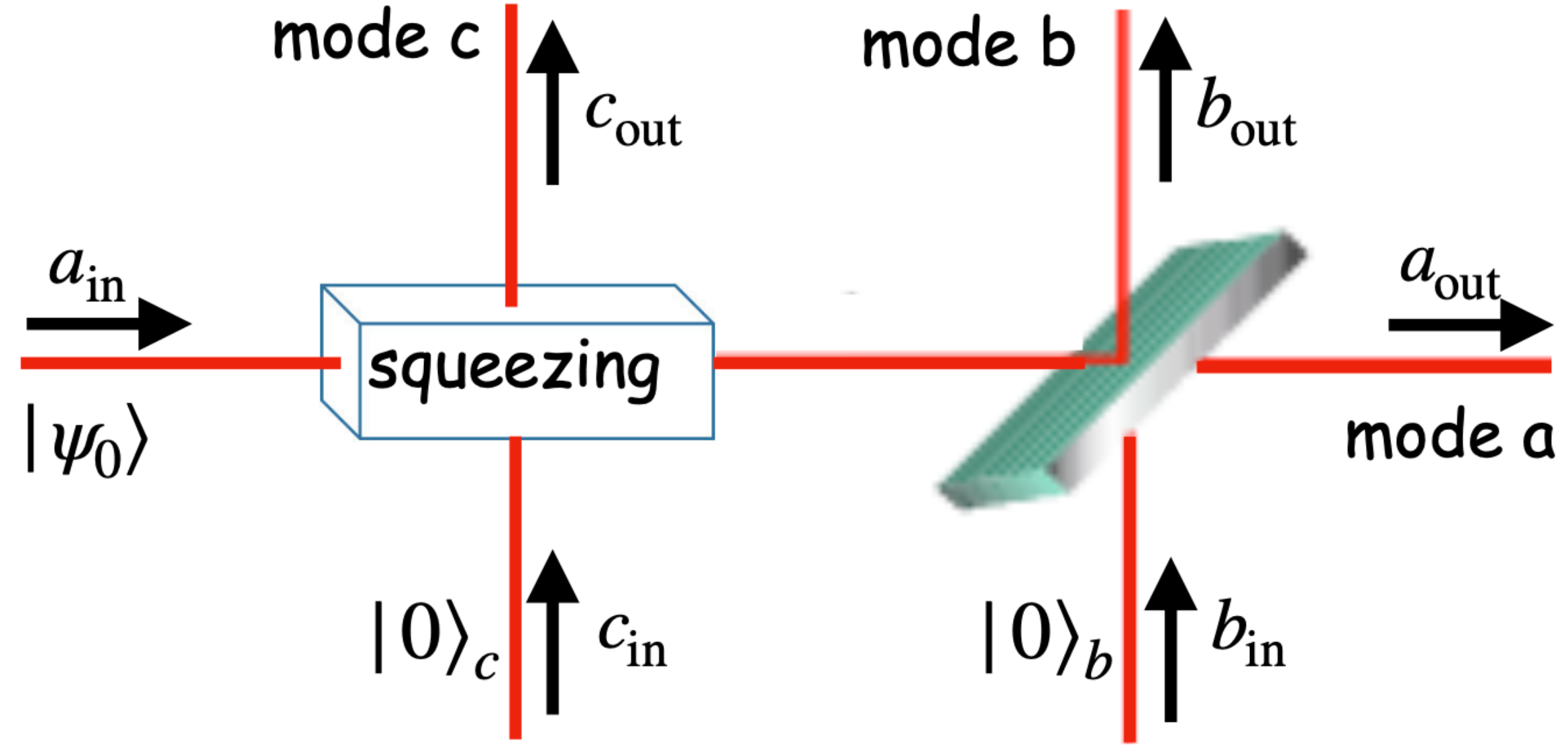}
\caption{Purification of the master equation for finite temperature by introducing two environments $b$ and $c$, initially in the vacuum state. The outgoing operators are obtained by applying a two-mode squeezing operation and a beam-splitter transformation to the incoming operators. Tracing out modes $b$ and $c$ recovers the master equation \eqref{master}.}
\label{Fig1}
\end{figure}

A possible purification of the corresponding evolution was derived in \cite{camille}. This is done by adding two independent environments, which can be represented by a beam-splitter and a two-mode squeezing operation, as shown in Fig.~\ref{Fig1}. 

We have then, with ${\cal B}$  and ${\cal S}$ corresponding respectively to the beam-splitter and squeezing transformations:
\begin{equation}\label{pure}
|\Psi(T)\rangle={\cal S} {\cal B}|\Psi_0\rangle|0\rangle_{b}|0\rangle_{c}\,,
\end{equation}
where
\begin{equation}
{\cal B}=\exp[\theta_1(ab^\dagger-a^\dagger b)]\,, \,\,
{\cal S}=\exp[\theta_2(a^\dagger c^\dagger-ac)]\,,
\end{equation}
and
\begin{eqnarray}
\theta_1(t)&=&\arccos\left[\sqrt{\eta\over n_T(1-\eta)+1}\right]\,,\\
\theta_2(t)&=&{\rm arccosh}\left[\sqrt{n_T(1-\eta)+1}\right]\,, \,\, \eta=e^{-2\gamma t}\,.
\end{eqnarray}
The corresponding operators are transformed as ${\cal O}_{\rm out}={\cal B}^\dagger{\cal S}^\dagger{\cal O}_{\rm in}{\cal S}{\cal B}$, as shown in Fig.~\ref{Fig1}.  We have then \cite{agarwal1}
\begin{equation}\label{aout}
a_{\rm out}=(a_{\rm in}\cos\theta_1-b_{\rm in}\sin\theta_1)\cosh\theta_2+c_{\rm in}^\dagger\sinh\theta_2\,.
\end{equation}

Equation \eqref{pure} leads to an upper bound to the QFI of the system. One should note that other purifications are possible. Indeed, addition of further unitary operations, depending only on the operators $b$ and $c$,  still lead to the same master equation. Variational parameters in these additional unitary transformations can be used to minimize the corresponding upper bound, so that it gets closer to the QFI of the system \cite{escher,camille,variational}. Here, however, we adopt the simpler procedure of using the purification \eqref{pure}, comparing the corresponding bound with the QFI of the open system.

For $T=0$, $\theta_2=0$ and ${\cal S}=1$, so mode $c$ gets decoupled from modes $a$ and $b$, implying that the corresponding master equation is purified with just a beam-splitter \cite{paris,adesso,agarwal1}, with transmission coefficient $\eta=\exp(-2\gamma t)$ and ${\cal B}=\exp\left[\phi(a^\dagger b-ab^\dagger)\right],\,\, \cos^2\phi\!=\!\eta$. From the corresponding generator $G(\gamma)=i[d{\cal B}^\dagger(\gamma)/d\gamma]{\cal B}(\gamma)$, one gets
\begin{equation}\label{lower}
\delta\gamma/\gamma\ge\delta\gamma_{\rm min}/\gamma={\left(e^{2\gamma t}-1\right)^{1/2}\over2\gamma t\bar{N}_{\rm in}^{1/2}}\,,
\end{equation}
where $\delta\gamma_{\rm min}$, obtained from $G(\gamma)$, is a lower bound for the uncertainty in the estimation of $\gamma$, and $t$ is the interaction time between the bosonic probe and the sample.

A simple way to estimate the standard deviation $\Delta\gamma$ is to use the error-propagation sensitivity expression $\Delta \gamma={\Delta N_{\rm out}/|\partial\bar{N}_{\rm out}/\partial\gamma|}$, where $\left(\Delta N_{\rm out}\right)^2=\langle(a_{\rm out}^\dagger a_{\rm out})^2\rangle-\bar N_{\rm out}^2$ is the variance in the boson distribution after the damping, and $\bar N=\langle a_{\rm out}^\dagger a_{\rm out}\rangle$ is the average number of bosons at the output. From \eqref{aout}, with $\theta_2=0$, one gets (the subscript $S$ stands for sensitivity):
\begin{equation}\label{sensitivity}
{\Delta \gamma_S\over\gamma}={[(\Delta N)^2_{\rm in}+(e^{2\gamma t}-1)\bar{N}_{\rm in}]^{1/2}\over2\gamma t\bar{N}_{\rm in}}
\xrightarrow[\Delta N\rightarrow0]{}\delta\gamma_{\rm min}/\gamma\,.
\end{equation}
This expression shows that the uncertainty in $\gamma$ has two contributions, the term $(\Delta N)^2_{\rm in}$ stemming from the initial variance in the bosonic number of the incoming probe, and the remaining terms corresponding to the random transmission of the incoming bosons. It is clear that, in order to minimize \eqref{sensitivity}, one must have $(\Delta N)^2_{\rm in}=0$, implying that the incoming bosons should be in a Fock state. In this case, $\Delta\gamma_S/\gamma$ becomes identical to the lower bound in \eqref{lower}! The presence of $\bar{N}^{1/2}_{\rm in}$ --  where $\bar{N}_{\rm in}$ is now just the number of bosons in the Fock state -- in the denominator of the right-hand side of \eqref{lower} implies that the same result would be obtained with a stream of $N$ independent single bosons. We note that $\delta\gamma\rightarrow\infty$ when $t\rightarrow0$ or $t\rightarrow\infty$, corresponding respectively to no action of the damping and to complete absorption, leading to no information on $\gamma$ (quantum Fisher information equal to zero). The minimum value of \eqref{lower} is obtained for
\begin{equation}\label{opt}
\gamma t_{\rm opt}\approx 0.8 \Rightarrow\delta\gamma_{\rm min}^{\rm opt}/\gamma=1.24/\sqrt{\bar{N}_{\rm in}}.
\end{equation}
This defines the optimal interaction time. Better precision can be obtained, however, by adopting a ``divide and conquer'' strategy. Instead of estimating the damping through a single measurement for an interaction time $t$, one applies sequential measurements, for instance with a single-photon stream, such that $t$ is divided into $N$ intervals of length $\tau$, which could be taken as the interaction time between each single photon and the probed sample. We replace then, in the right-hand side of \eqref{lower}, $t$ by $\tau$ and $\bar{N}_{\rm in}$ by $N=t/\tau$. The corresponding expression is minimized for $\tau\rightarrow0$. However, any other $\tau$ smaller than $t$ would lead to a result better than measuring just at time $t$. For $\gamma\tau\ll1$, $\delta\gamma/\gamma\approx1/\sqrt{2\gamma t}$, which is much smaller than \eqref{opt} if $\gamma t\gg1$. We note that this strategy not only leads to better precision, but could be mandatory for thin or fragile samples, for which the interaction time with the probe should necessarily be smaller than the thermalization time.

Confirmation of this result is obtained by explicitly calculating the quantum Fisher information for incoming Fock states. The general expression for the quantum Fisher information for estimation of a parameter $X$ is expressed in terms of the density operator of the probe as ${\cal F}_Q(X)={\rm Tr}\left[\rho(X)L^2(X)\right]$, where the {\it symmetric logarithmic derivative} is defined by the equation $d\rho(X)/dX=[\rho(X)L(X)+L(X)\rho(X)]/2$. Finding $L$ requires, in general, the diagonalization of the density operator, for a given initial state \cite{caves,paris}. However, for incoming Fock states the density matrix is diagonal, and therefore the singular logarithmic derivative is given by $L_{nn}=(1/p_n)(dp_n/dX)$, where $p_n\equiv\rho_{nn}$ is the boson-number probability distribution.  It follows then that
\begin{equation}\label{fisher}
{\cal F}_Q(\gamma)={\rm Tr}\left(\rho{L}^2\right)=\sum_n(1/ p_n)(dp_n/ d \gamma)^2\,,
\end{equation}
coinciding with the Fisher information associated to the measurement of the bosonic population distribution, which is thus shown to be the best measurement in this case.  On the other hand, the boson-number distribution for the outgoing bosons is identical to the beam-splitter binomial distribution, $p_n(\gamma)={N\choose n}(1-\eta)^{N-n}\eta^n$. Replacing this expression in \eqref{fisher} leads precisely to \eqref{lower}. Furthermore, as $N\gg1$ (which could apply to a Fock state or a stream of single photons), the combinatorial distribution goes to a Gaussian distribution, with width given by the lower bound in \eqref{lower}, so this bound is actually saturated by these states. This completes our demonstration that Fock states lead to the minimal uncertainty in the estimation of $\gamma$ \cite{added1}.

For $T\not=0$, one gets a lower bound $\Delta\gamma^{G}_{\rm min}(T)$ from the unitary transformation in \eqref{pure} (details in supplementary material sec I):
\begin{align}\label{lowerT}
\!\!&\delta\gamma^{G}_{\rm min}(T)/\delta\gamma_{\rm min}\nonumber\\
\!\!&={n_T(1-\eta)+1\over \sqrt{n_T(1+\eta^2)+1+(n_T/\bar{N}_{\rm in})\eta[n_T(1-\eta)+1]}}\,,
\end{align}

where $\delta\gamma_{\rm min}$ is defined in \eqref{lower}. Calculations also show that \eqref{lowerT}, for any T, is lower than the bound calculated using error propagation sensitivity (see supplementary material Sec II).

\begin{figure}[tb]
\includegraphics[width=\columnwidth]{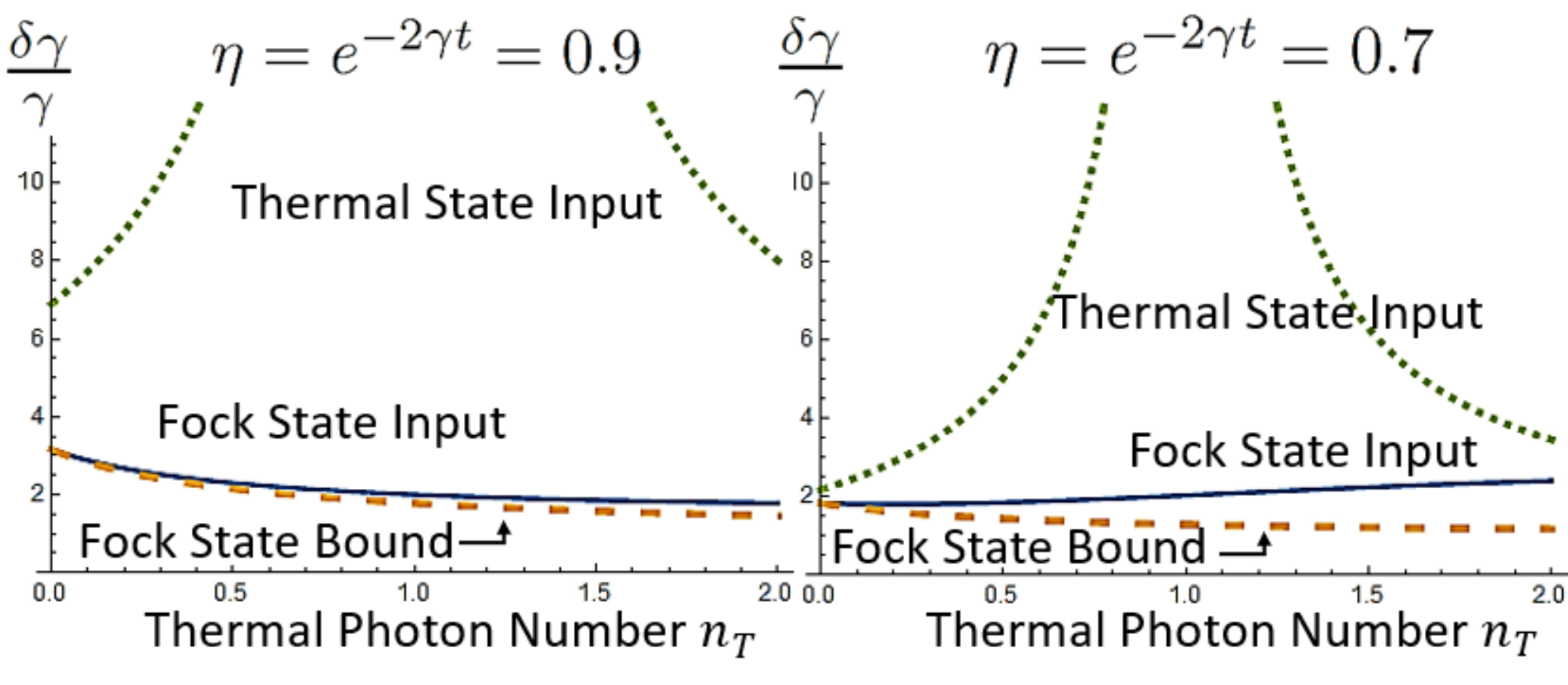}
\caption{$\delta\gamma/\gamma$ as a function of the bath thermal photon number, for two different values of $\eta\equiv\exp(-2\gamma t)$. The solid and dotted curves correspond to single-photon and thermal state inputs,  this last one with an average photon number equal to one. The dashed curve corresponds to the bound \eqref{lowerT}, obtained from the purification procedure. For $\eta=0.9$, and single-photon input, precision increases with temperature, for the range here considered.}
\label{Fig2}
\end{figure}

The QFI of the system, for incoming Fock states, can be calculated numerically, from the number probability distribution given in \cite{agarwal,me} -- see Eq.~(S20) in the Supplementary Material. It coincides with \eqref{lowerT} when there is no input i.e. $\bar{N}_{\rm in}=0$. In this case, only thermal photons contribute to the estimation of $\gamma$ (supplementary material sec III). Fig.~\ref{Fig2} shows the behavior of $\delta\gamma/\gamma$ for $\bar{N}_{\rm in}=1$ and two values of $\eta=\exp{(-2\gamma t)}$, namely $\eta=0.9$ and $\eta=0.7$.  As expected, say from \eqref{sensitivity}, the incoming thermal state is a poor choice for estimation of $\gamma$. In case of initial thermal state with $\bar{N}_{\rm in}=n_T$, there is no time evolution of the incoming state, and hence the quantum Fisher information vanishes, which leads to the divergent behavior of the dotted curve in Fig.~\ref{Fig2}.

\begin{figure}[tb]
\includegraphics[width=\columnwidth]{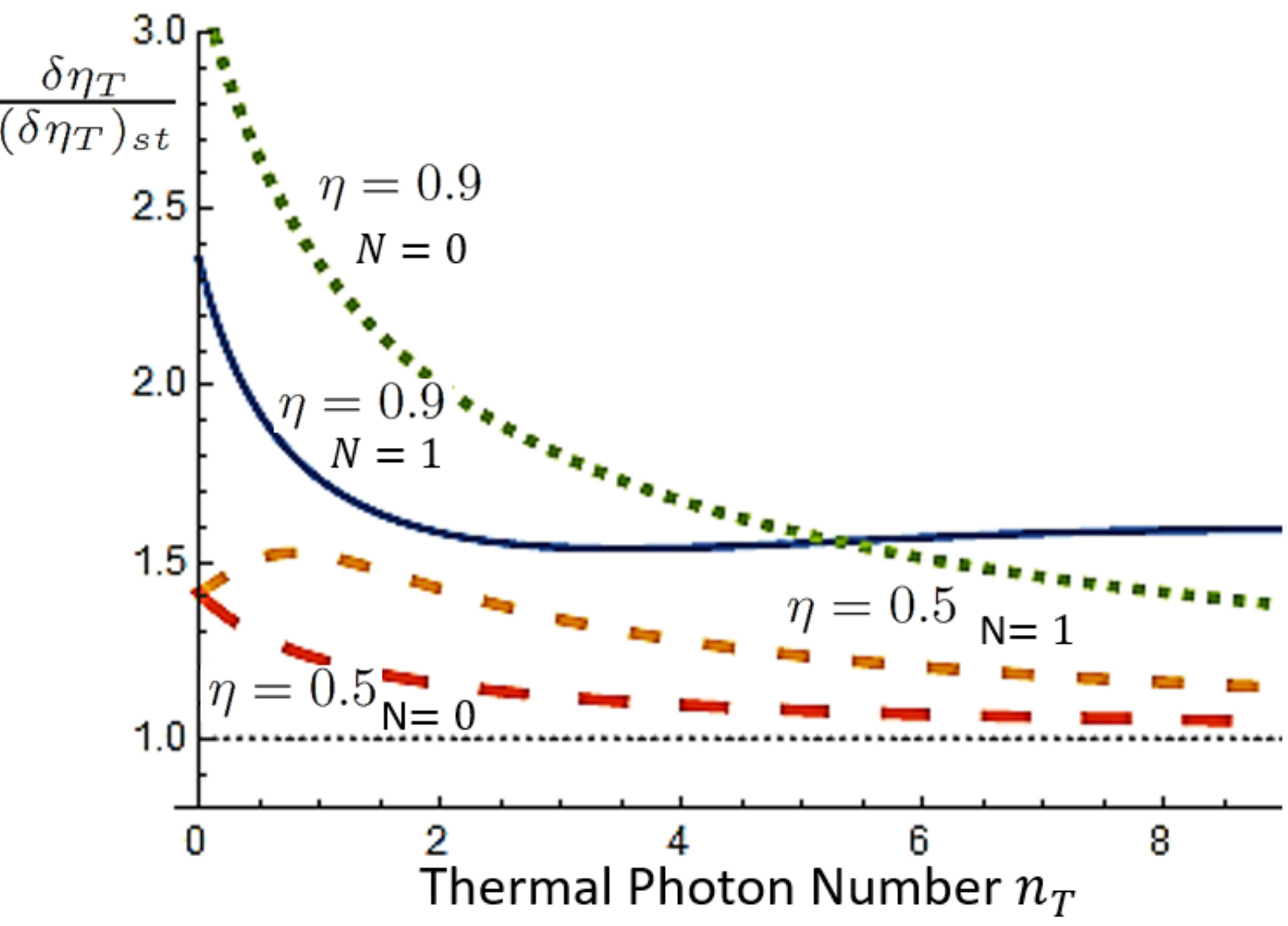}
\caption{Uncertainty $\delta n_T$ in the measurement of temperature, normalized by the steady state value $(\delta n_T)_{\rm st}$, for different values of $\eta=\exp(-2\gamma t)$. Each curve is labelled by the photon number $N$ in the incoming Fock state of the probe. In the limit $t\rightarrow\infty$, so that $\eta\rightarrow0$, one has $\delta\eta_T/(\delta\eta_T)_{st}=1$. The graph suggests that the best measurement occurs for large $t$ (or small  $\eta$). For  $\eta=0.9$ ($\gamma t\sim0.1$), and for $n_T\le 4.5$, single-boson Fock state leads to better precision than the vacuum state $|0\rangle$. Sequential measurements may lead however to much better precision, as shown in the text of the article.}
\label{Fig3}
\end{figure}

\noindent{\it Estimation of temperature}. The simplest situation corresponds to no incoming photons. In this case, the beam splitter in \eqref{pure} does not play a role, and the purification is given by $|\Psi(t)\rangle={\cal S}|0\rangle_S|0\rangle_{R_{1}}|0\rangle_{R_2}$. From the generator $G(n_T)=i[d{\cal S}^\dagger(n_T)/dn_T]{\cal S}(n_T)$, one gets then an upper bound for the quantum Fisher information, from which it follows a lower bound for the uncertainty in the estimation of $n_T$:
\begin{equation}\label{temp}
\delta n_T=\sqrt{n_T\!\left(n_T\!+{1\over1-\eta}\right)}\xrightarrow[t\rightarrow\infty]{}\sqrt{n_T(n_T+1)}.
\end{equation}
For no incoming photon the sensitivity expression and the QFI yield for $\delta n_T$ the same result. Therefore, in this case the lower bound for the uncertainty coincides with the exact result.  As the interaction time between probe and sample increases, $\delta n_T$ is reduced, attaining the steady-state limit $(\delta n_T)_{st}=\sqrt{n_T(n_T+1)}$ when $t\rightarrow\infty$, which coincides with the quantum-mechanical uncertainty for a thermal field.  The numerical results from the solution of the master equation for an incoming single photon state are shown in Fig.~\ref{Fig3}. The Fock state $|1\rangle$ leads to better precision for small times and low temperatures, as compared the vacuum state $|0\rangle$.

As in the estimation of damping, an increase in precision can be obtained by applying sequential  pre-thermalization measurements, through a stream of single bosons. The measurement time $t$ is divided into $\nu$ intervals of length $\tau$, corresponding to the interaction between a single boson and the probed system. The corresponding QFI ${\cal F}_Q(n_T,\tau)$ can be obtained from \eqref{master} in the small-time limit $\gamma\tau (n_T+1)\ll1$, and the corresponding uncertainty is $\delta n_T=1/\sqrt{(t/\tau){\cal F}_Q(n_T,\tau)}$. It turns out that the best result is obtained when $\tau\rightarrow0$, but any other $\tau$ smaller than $t$ would lead to a better result then measuring at $t$.  In the limit $\gamma\tau (n_T+1)\ll1$, we get (supplementary material Sec IV)
\begin{equation}
\delta n_T\rightarrow\sqrt{{n_T(n_T+1)\over (3n_T+2)2\gamma t}}\xrightarrow[n_T\ll1]{}\sqrt{n_T/4\gamma t}\,.
\end{equation}
When $\gamma t\gg 1$, this expression is much smaller than $(\delta n_T)_{st}$, implying a huge gain in precision, as compared to measurement at time $t$. The effect on the protocol by timing errors can be easily accounted for, since the above expression depends only on the total time $t$. For $\Delta t/t\ll1$, then the extra uncertainty in the temperature estimation, $\Delta(\delta n_T)$, will be much smaller than $\delta n_T$.  

\par
\par\noindent{\it Conclusion}. We have established the quantum precision limits for the estimation of damping constants and temperature, when bosons are used as probes. Bosonic probes occupy a prominent place in science, especially in view of the large number of processes involving light or microwave fields to obtain information on absorption coefficients or the temperature of transparent samples. Lower analytic bounds for the uncertainty in the estimation of these parameters have been obtained, through a purification procedure that involves replacing the master equation by a unitary transformation composed by a beam splitter and a squeezing operator, acting on the bosonic mode and two auxiliary environments. These bounds were shown to be tight, for some specific conditions, and, more generally, close to the numerical solutions. We have shown that sequential pre-thermalization measurements with single-photon streams can lead to huge gain in precision, both for damping and temperature estimation.  This result is especially relevant for measurements on thin or fragile samples. We believe these findings should stimulate experimental work on physical and biological systems.

LD acknowledges the support of the Brazilian agencies CNPq, CAPES, FAPERJ, of the National Institute of Science and Technology for Quantum Information, and of the Hagler Institute for Advanced Study of the Texas A\&M University. G.S.A. thanks the support of Air Force Office of Scientific Research (Award N\textsuperscript{\underline{o}} FA-9550-18-1-0141) and the Robert A
Welch Foundation (A-1943-20180324).
\section{SUPPLEMENTARY MATERIAL}

\subsection{I. Lower bounds on uncertainties in the estimation of damping and temperature}

Here we provide the derivation of lower bounds on the uncertainties in the estimation of damping and temperature by using the purification procedure described in the text, corresponding to Fig.~1. If the probe is in the initial state $|\Psi_0\rangle $, and interacts with the probed system during a time $t$, then the purified output state is 
\begin{equation}
|\Psi(t)\rangle={\cal S} {\cal B}|\Psi_0\rangle|0\rangle_{b}|0\rangle_{c}\,,\tag{${S1}$}
\end{equation}
where the two environments $b$ and $c$ are assumed to be initially in the vacuum state. 
\noindent The operators ${\cal S}$ and ${\cal B}$ are defined by Eqs.~(3)-(5)  in the main text. The operator $G(X)\equiv i[dU^\dagger(X)/dX]U(X)$, for an arbitrary parameter $X$, where 
$U(X)={\cal S}(X){\cal B}(X)$, is given by 

\begin{align}
\label{S2}
G(X)=& -i\{(ab^{\dagger}-a^{\dagger}b)\frac{d\theta_{1}}{dX}+[c^{\dagger}(a^{\dagger}\cos\theta_{1}-b^{\dagger}\sin\theta_{1})\nonumber\\
& -c(a\cos\theta_{1}-b \sin\theta_{1})]\frac{d\theta_{2}}{dX}\}\,,
\tag{${S2}$}
\end{align}

where $\theta_1$ and $\theta_2$ are defined by Eqs.~(4) and (5) in the main text,  and $a$, $b$, and $c$ are annihilation operators corresponding respectively to the original bosonic mode and the additional environments $b$ and $c$.
\noindent On applying this operator to the initial state $|\Psi_0\rangle \left|0\right\rangle_{b} \left|0\right\rangle_{c} $, one gets: 

\begin{align}
\label{S3}
&G|\Psi_0\rangle |0 \rangle_{b} |0\rangle_{c} =-i[a|\Psi_0\rangle |1\rangle_{b} |0\rangle_{c} \frac{d\theta_{1}}{dX}\nonumber\\
&+(a^{\dagger}|\Psi_0\rangle |0\rangle_{b} |1\rangle_{c} \cos\theta_{1}-|\Psi_0\rangle |1\rangle_{b} |1\rangle_{c} \sin\theta_{1})\frac{d\theta_{2}}{dX}]\,.
\tag{${S3}$}
\end{align}

\noindent The expectation value of the operator $G^{\dagger}G$ in the state $|\Psi_0\rangle \left|0\right\rangle_{b} \left|0\right\rangle_{c} $ is therefore
\begin{equation}
\left\langle G^{\dagger}(X)G(X)\right\rangle =\overline{N}_{\rm in} \left[(\frac{d\theta_{1}}{dX})^{2}+\cos^{2}\theta_{1}(\frac{d\theta_{2}}{dX})^{2}\right]+(\frac{d\theta_{2}}{dX})^{2},\tag{${S4}$}
\end{equation}
where $\overline{N}_{\rm in}=\left\langle a^{\dagger}a\right\rangle $ is the average number of photons in the input state $|\Psi_0\rangle$. After simplification we get
the final expressions for $X=\gamma$ and $X=n_{T}$, where $\gamma$ is the damping coefficient and $n_T$ is the thermal photon number, as functions of $\eta$, $n_{T}$ and $\overline{N}_{\rm in}$:
\begin{align}
\left\langle G^{\dagger}(\gamma)G(\gamma)\right\rangle &=\frac{(2t\eta)^{2}}{4}\{\overline{N}_{\rm in}\frac{1+n_{T}(1+\eta^{2})}{\eta(1-\eta)[1+n_{T}(1-\eta)]^{2}}\nonumber\\
&+\frac{n_{T}}{(1-\eta)[1+n_{T}(1-\eta)]}\}\,;
\tag{${S5}$}
\end{align}

\begin{align}
&\left\langle G^{\dagger}(n_{T})G(n_{T})\right\rangle =\frac{1}{4}\{\frac{(1-\eta)}{[1+n_{T}(1-\eta)]n_{T}}\nonumber\\
&+\overline{N}_{\rm in}\frac{\eta^{2}(1-\eta)\{n_{T}(1-\eta)[1+n_{T}]+(1+n_{T})\}}{[1+n_{T}(1-\eta)]^{3}(1+n_{T})n_{T}}\}\,.\tag{${S6}$}
\end{align}

\noindent Since the Quantum Fisher Information ${\cal F}_Q=4(\Delta G)^2=\left\langle G^{\dagger}(n_{T})G(n_{T})\right\rangle-\langle G\rangle^2$,  the lower bound for the uncertainty in the estimation of damping and temperature can be calculated from  $\delta X_{\rm min}^{G}=\mathit{\mathscr{\mathcal{F}}}_{Q}^{\,-1/2}$:
\begin{align}
&\delta\gamma{}_{\rm min}^{G}=\frac{1+n_{T}(1-\eta)}{2t\eta}[\eta(1-\eta)]^{1/2}\nonumber\\
&\times\{\overline{N}_{\rm in}[1+n_{T}(1+\eta^{2})]+\eta[1+n_{T}(1-\eta)]n_{T}\}^{-1/2}\,;
\tag{${S7}$}
\end{align}

\begin{align}
&\delta n_{T}{}_{\rm min}^{G}=\{[1+n_{T}(1-\eta)]^{3}(1+n_{T})n_{T}\}^{1/2}\nonumber\\
&\times\{\overline{N}_{\rm in}\eta^{2}[n_{T}(1-\eta)(2+n_{T}-\eta n_{T}))+(1-\eta)]\nonumber\\
&+[1+n_{T}(1-\eta)]^{2}(1+n_{T})(1-\eta)\}^{-1/2}\,.
\tag{${S8}$}
\end{align}

\noindent Comparing (S7) with the one at zero temperature,
$\delta\gamma_{\rm min}=\frac{\sqrt{1/\eta-1}}{2t\sqrt{\overline{N}_{\rm in}}}$, we get Eq.~(11) in the main text,
\begin{align}
\frac{\delta\gamma{}_{\rm min}^{G}}{\delta\gamma_{\rm min}}&=[1+n_{T}(1-\eta)]\{[1+n_{T}(1+\eta^{2})]\nonumber\\
&+\eta[1+n_{T}(1-\eta)]n_{T}/\overline{N}_{\rm in}\}^{-1/2}\,.\tag{${S9}$}
\end{align}

\noindent For vacum input, the expression for $\delta n_{T}{}_{\rm min}^{G}$
 becomes Eq.(12) in the main text,
\begin{equation}
\delta n_{T}=\sqrt{\frac{n_{T}[1+n_{T}(1-\eta)]}{1-\eta}}.\tag{${S10}$}
\end{equation}

\subsection{II. Sensitivity calculations using the master equation}

The error-propagation expression for the uncertainty in the estimation of a parameter $X$ is given by $\Delta X= \Delta N_{\rm out}/\frac{\partial \overline{N}_{\rm out}}{\partial X}$, where $(\Delta N_{\rm out})^2=\langle(a^\dagger_{\rm out}a_{\rm out})^2\rangle-\overline{N}_{\rm out}^2$. From Eq.~(1) in the main text, we can study the evolution of an operator $A$ by $\frac{\partial}{\partial t}\overline{A}=Tr[\frac{\partial\rho_{s}}{\partial t}A]$. For any operator $A$,

\begin{align}
\frac{\partial}{\partial t}\overline{A}&=-\gamma(1+n_{T})<Aa^{\dagger}a-2aAa^{\dagger}+a^{\dagger}aA>\nonumber\\
&-\gamma n_{T}<aa^{\dagger}A-2a^{\dagger}Aa+Aaa^{\dagger}>\,.\tag{${S11}$}
\end{align}

\noindent Since $Tr[AB]=Tr[BA],$ we have

\begin{align}
\frac{\partial}{\partial t}\overline{A}&=-\gamma(1+n_{T})<[A,a^{\dagger}]a+a^{\dagger}[a,A]>\nonumber\\
&-\gamma n_{T}<[A,a]a^{\dagger}+a[a^{\dagger},A]>\,.\tag{${S12}$}
\end{align}

\noindent Taking $A=(a^{\dagger}a)^{i}\,,\:(i=1,2)$, one gets
\begin{equation}
\frac{\partial}{\partial t}\overline{N}=-2\gamma\overline{N}+2\gamma n_{T},\tag{${S13}$}
\end{equation}
\begin{equation}
\frac{\partial}{\partial t}\overline{N^{2}}=-4\gamma\overline{N^{2}}+2\gamma(4n_{T}+1)\overline{N}+2\gamma n_{T}.\tag{${S14}$}
\end{equation}

\noindent Integrating these equations on both sides one gets
\begin{equation}
\overline{N}(t)=e^{-2\gamma t}(\overline{N}(0)-n_{T})+n_{T},\tag{${S15}$}
\end{equation}
\begin{align}
\overline{N^{2}}(t)&=e^{-4\gamma t}\overline{N^{2}}(0)+e^{-2\gamma t}(4n_{T}+1)(1-e^{-2\gamma t})\overline{N}(0)\nonumber\\
&+2n_{T}^{2}(1-e^{-2\gamma t})^{2}+n_{T}(1-e^{-2\gamma t})\,.\tag{${S16}$}
\end{align}

\noindent Note that $\overline{N}_{\rm out}=\overline{N}(t)$ and $\overline{N}_{\rm in}=\overline{N}(0)$. Using (S15) and (S16) we find

\begin{align}
&\triangle\gamma=[\eta^{2}(\triangle \overline{N}_{\rm in})^{2}+\eta(2n_{T}+1)(1-\eta)\overline{N}_{\rm in}\nonumber\\
&+(n_{T}+1-\eta n_{T})n_{T}(1-\eta)]^{1/2}[2t\eta (\overline{N}_{\rm in}-n_T)]^{-1}\,.\tag{${S17}$}
\end{align}
\noindent For given $\overline{N}_{\rm in}$, $\eta$, and $n_T$, the minimal uncertainty $\triangle\gamma_{\rm min}$ is achieved for $(\triangle \overline{N}_{\rm in})^{2}=0$, indicating that Fock states lead to the best estimation of $\gamma$.

We compare now (S17), for Fock states, so that $\overline{N}_{\rm in}=N_{\rm min}$, with the bound $\delta\gamma{}_{\rm min}^{G}$, obtained in Sec I using purification. The
ratio of $\triangle\gamma_{\rm min}$ and $\delta\gamma{}_{\rm min}^{G}$ .
\begin{align}
&\frac{\triangle\gamma_{\rm min}}{\delta\gamma{}_{\rm min}^{G}}=(n_{T}(1-\eta)+1)\frac{\overline{N}_{\rm in}}{\overline{N}_{\rm in}-n_{T}}\nonumber\\
&\times\sqrt{\frac{N_{\rm in}\eta(2n_{T}+1)+n_{T}[n_{T}(1-\eta)+1]}{N_{\rm in}\eta[n_{T}(1+\eta^{2})+1]+\eta n_{T}[n_{T}(1-\eta)+1]}}\,.\tag{${S18}$}
\end{align}

\noindent Since $n_{T}(1-\eta)+1\geq1$, $N_{\rm in}\eta(2n_{T}+1)\geq N_{\rm in}\eta[n_{T}(1+\eta^{2})+1]$, and $n_{T}[n_{T}(1-\eta)+1]\geq\eta n_{T}[n_{T}(1-\eta)+1]$, one gets 
\begin{equation}
\triangle\gamma_{\rm min}\geq\delta\gamma{}_{\rm min}^{G}.\tag{${S19}$}
\end{equation}

\noindent The equality sign in (S19) holds only when $n_{T}=0$, which coincides with the discussion after Eq.~(8) in the main text: at zero temperature, the error-propagation formula for the  estimation uncertainty coincides with the lower bound. It may be noted that expressions like (S17) are not meaningful when $\overline{N}_{\rm in}$ approaches $n_{T}$. In this limit the output photon number becomes independent of the parameter $\gamma$, which we had set out to determine. In such cases post processing of signal is needed. It may be added that the full master equation solution for the input Fock state has no such divergence as the bound is calculated using full photon number distribution. For thermal input with input photon number equal to $n_T$, master equation solution gives divergence [Fig.2] because as noted there the Fisher information becomes zero and not meaningful as the system does not evolve then.

\subsection{III. Master equation result for the QFI with no incoming photons}

The solution of the master equation given by Eq.~(1) was studied numerically in the paper for both Fock states and thermal
states. However it is possible to get the analytical result for vacuum
input. From \cite{agarwal} we get the probability of seeing $n$ photons at
the output state with input Fock state $|m\rangle$:
\begin{align}
p_{n}&=\frac{(1-e^{-2\gamma t})^{n+m}(e^{\beta\omega}-1)e^{m\beta\omega}}{(e^{\beta\omega}-e^{-2\gamma t})^{n+m+1}}\nonumber\\
&\times F[-n,-m,1:\frac{e^{\beta\omega}+e^{-\beta\omega}-2}{e^{2\gamma t}+e^{-2\gamma t}-2}]\,,\tag{${S20}$}
\end{align}

\noindent where $e^{\beta\omega}=1+n_{T}^{-1}$ and $F$ is the hypergeometric function. 
For $m=0$, $F[-n,0,1:z]=1$, thus we have $p_{n}=\frac{(1-\eta)^{n}n_{T}^{-1}}{(1+n_{T}^{-1}-\eta)^{n+1}}$. Let $\overline{n}(t)=n_{T}(1-\eta)$, then $p_{n}$ can be written as the Bose-Einstein distribution

\begin{equation}
p_{n}=\frac{\overline{n}(t)^{n}}{(1+\overline{n}(t))^{n+1}}.\tag{${S21}$}
\end{equation}

\noindent From (S21) we obtain the Quantum Fisher Information
\begin{equation}
F_{Q}=\frac{n_{T}^{2}(2t\eta)^{2}}{\overline{n}(t)(1+\overline{n}(t))}.\tag{${S22}$}
\end{equation}

\noindent With $\delta\gamma\geq\sqrt{{\mathit{\mathscr{\mathcal{F}}}_{Q}}^{-1}}$, we get the lower bound for
$\gamma$ as 

\begin{equation}
\delta\gamma^{G}(T)=\frac{1}{2t\eta}\sqrt{\frac{(1-\eta)[1+n_{T}(1-\eta)]}{n_{T}}}.\tag{${S23}$}
\end{equation}

\noindent This coincides with $\delta\gamma{}_{\rm min}^{G}$ in (S7) with $N_{\rm in}=0$ obtained in section
I with purification.

\subsection{IV. Estimation of bounds for the uncertainty in temperature estimation with a stream of single photons}

We consider now the bound for the uncertainty in   $\delta n_{T}$
with a stream of single photons, each one interacting with the probed system  during a time interval of $\tau$, so that the total interaction time is divided into $\nu$ intervals, with $t=\nu\tau$. From the master equation, we get the dynamics of $p_{n}$, the probability of detecting $n$ bosons, after they have interacted with the sample:
\begin{align}
\frac{dp_{n}}{dt}&=2\gamma(n_{T}+1)[(n+1)p_{n+1}-np_{n}]\nonumber\\
&+2\gamma n_{T}[np_{n-1}-(n+1)p_{n}]\,.\tag{${S24}$}
\end{align}

\noindent Here if we have a single-boson input at each time interval $\tau$, so $p_{n}(0)=\delta_{n,1}$. We integrate equation (S23) assuming $\gamma\tau (n_{T}+1)\ll1$, so that
\begin{align}
p_{n}(\tau)&\simeq\delta_{n,1}+2\gamma\tau(n_{T}+1)[(n+1)p_{n+1}-np_{n}]\nonumber\\
&+2\gamma\tau n_{T}[np_{n-1}-(n+1)p_{n}]\,,\tag{${S25}$}
\end{align}

\noindent where $p_{n}$ on the right side gives the distribution at $\tau=0$. From (S25) we then obtain
\begin{equation}
\begin{array}{c}
p_{0}(\tau)=2\gamma\tau(n_{T}+1),\\
p_{1}(\tau)=1-2\gamma\tau(n_{T}+1)-4\gamma\tau n_{T},\\
p_{2}(\tau)=4\gamma\tau n_{T}.
\end{array}
\tag{${S26}$}
\end{equation}

\noindent For a total interaction time $t=\nu\tau$, corresponding to $\nu=t/\tau$ single-boson interactions, the Quantum Fisher Information is then
\begin{figure}[tb]

\includegraphics[width=\columnwidth]{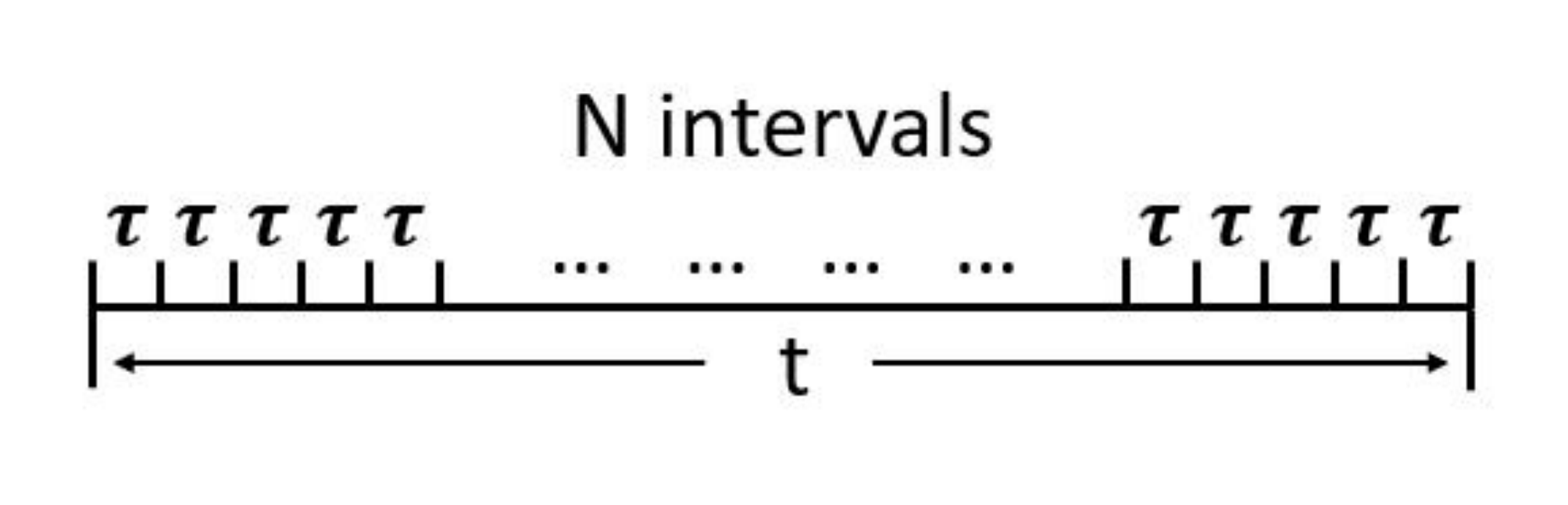}
\caption{A huge increase in the estimation precision can be obtained with a stream of $N$ single bosons, each one interacting with the probed material for a time $\tau$ much smaller than the thermalization time $t$.}
\label{S1}
\end{figure}
\begin{align}
&\mathit{\mathscr{\mathcal{F}}}_{Q}(t)\!=\!\nu\mathit{\mathscr{\mathcal{F}}}_{Q}(\tau)\!=\!\nu\sum_n{1\over p_n}\left({dp_n\over dn_T}\right)^2\!\nonumber\\
&=\!\frac{t}{\tau}[\frac{(2\gamma\tau)^{2}}{2\gamma\tau(n_{T}+1)}+\frac{(6\gamma\tau)^{2}}{1-2\gamma\tau(n_{T}+1)-4\gamma\tau n_{T}}+\frac{(4\gamma\tau)^{2}}{4\gamma\tau n_{T}}]\nonumber\\
&\overset{\tau\shortrightarrow0}{\longrightarrow}2\gamma t\frac{3n_{T}+2}{n_{T}(n_{T}+1)}\,.
\tag{${S27}$}
\end{align}

\noindent The corresponding lower bound for the uncertainty in the estimation of the thermal photon number  is obtained from $\delta n_{T}=1/\sqrt{\frac{t}{\tau}\mathit{\mathscr{\mathcal{F}}}_{Q}(\tau)}$:
\begin{equation}
\delta n_{T}\overset{\tau\shortrightarrow0}{\longrightarrow}\sqrt{\frac{n_{T}(n_{T}+1)}{(3n_{T}+2)2\gamma t}}.\tag{${S28}$}
\end{equation}
\noindent In the low temperature limit $n_{T}\ll1$, we have
\begin{equation}
\delta n_{T}|_{n_{T}\ll1}\overset{\tau\shortrightarrow0}{\longrightarrow}\sqrt{\frac{n_{T}}{4\gamma t}}.\tag{${S29}$}
\end{equation}

\noindent The limits (S28), (S29) are discussed in the main text.


\begin{thebibliography}{99}{
\bibitem{cramer} H. Cram\'er, ``Mathematical Methods of Statistics,'' p. 500, Princeton University, Princeton, NJ, USA (1946).
\bibitem{fisher} R. A. Fisher, Proc. R. Soc. Edinburgh {\bf 42}, 321 (1922).
\bibitem{helstrom} C. W. Helstrom, ``Quantum Detection and Estimation Theory,'' Chap. VIII.4, Academic Press,
New York (1976).
\bibitem{holevo} A. S. Holevo, ``Probabilistic and Statistical Aspects of Quantum Theory,'' North-Holland,
Amsterdam (1982).
\bibitem{caves} C. M. Caves, Phys. Rev. D {\bf 23}, 1693 (1981).
\bibitem{caves2} S. L. Braunstein and C. M. Caves, Phys. Rev. Lett. {\bf 72}, 3439 (1994).
\bibitem{gravitation} H. Grote, K. Danzmann, K. L. Dooley, R. Schnabel, J. Slutsky, and H. Vahlbruch, Phys. Rev. Lett. {\bf 110}, 181101 (2013).
\bibitem{schnabel} J. S\"udbeck, S. Steinlechner, M. Korobko, and R. Schnabel, Nat. Photonics 1-4 (2020).
\bibitem{sewell} R. J. Sewell {\it et al}, Phys. Rev. Lett. {\bf 109}, 253605 (2012).
\bibitem{thermo} L. A. Correa, M. Mehboudi, G. Adesso, and A. Sanpera, Phys. Rev. Lett. {\bf 114}, 220405 (2015).
\bibitem{facon} A. Facon {\it et al}, Nature (London){\bf  535}, 262 (2016).
\bibitem{bowen} M. A. Taylor {\it et al}, Phys. Rev. X {\bf 4}, 011017 (2014).
\bibitem{taylor} M. A. Taylor and W. P. Bowen, Phys. Rep. {\bf 615}, 1 (2016).
\bibitem{gefen} T. Gefen, A. Rotem, and A. Retzker, Nat. Commun. {\bf 10}, 4992 (2019).
\bibitem{sensing} C. L. Degen, R. Reinhard, and P. Cappellaro, Rev. Mod. Phys. {\bf 89}, 035002 (2017).
\bibitem{pezze} L. Pezz\`e {\it et al}, Rev. Mod. Phys. {\bf 90}, 035005 (2018).
\bibitem{diamond} D. Kim {\it et al}, Nat. Electron. {\bf 2}, 284 (2019).
\bibitem{nano}  C. Li,  M, Chen, D. Lyzwa, and P. Cappellaro, Nano Lett. {\bf 19}, 7342 (2019).
\bibitem{bouton} Q. Bouton {\it et al}, Phys. Rev. X {\bf 10}, 011018 (2020).
\bibitem{multi1} M. Szczykulska, T. Baumgratz, and A. Datta, Adv. Phys. X 1, 621 (2016).
\bibitem{multi2} F. Albarelli, J. F. Friel, and A. Datta, Phys. Rev. Lett. {\bf 123}, 200503 (2019).
\bibitem{henning} P. A. Henning, Condensed Matter Physics {\bf 3}, 75 (2000).
\bibitem{jenson} J. R. Jenson, ``Remote Sensing of the Environment: An Earth Resource Perspective,'' Person Prentice Hall, NJ, USA (2007).
\bibitem{haswell} S. J. Haswell,  ``Atomic Absorption Spectrometry; Theory, Design and Applications,'' Elsevier, Amsterdam (1991).
\bibitem{wit} J. de Wit, S. Seager, Science, {\bf 342}, 1473 (2013).
\bibitem{biology} M. T. Cone {\it et al}, Optica {\bf 2}, 162 (2015).
\bibitem{thermometry} W. Weng {\it et al}, Phys. Rev. Lett. {\bf 112}, 160801 (2014).
\bibitem{NIST} H. W. Yoon, V. Khromchenko, and G. P. Eppeldauer, Optics Express {\bf 27}, 14246 (2019).
\bibitem{paris} A. Monras and M. G. Paris, Phys. Rev. Lett. {\bf 98}, 160401 (2007).
\bibitem{braun} P. Binder and D. Braun, ArXiv:1905.08288v1 [quant-ph] (2019).
\bibitem{adesso} G. Adesso {\it et al}, Phys. Rev. A {\bf 79}, 040305 (2009).
\bibitem{escher} B. M. Escher, R. L. de Matos Filho, and L. Davidovich, Nature Phys. {\bf 7}, 406 (2011);  Braz. J. Phys. {\bf 41}, 229, (2011).
\bibitem{camille} C. L. Latune, B. M. Escher, R. L. de Matos Filho, and L. Davidovich, Phys. Rev. A {\bf 88}, 042112 (2013).
\bibitem{variational} B. M. Escher, L. Davidovich, N. Zagury, and R. L. de Matos Filho, Phys. Rev. Lett. {\bf 109}, 190404 (2012).
\bibitem{agarwal1} G. S. Agarwal, {\it Quantum Optics}, Cambridge University Press, Secs 5.1 and 5.10 (2013).
\bibitem{added1} It can be seen from the results of \cite{paris} , that ratio $\delta\gamma/\delta\gamma_{\rm min}$ is higher by $1\over \sqrt{1-\eta}$ for coherent state input and by ${\{1+\frac{2\eta(1-\eta)}{2\eta^2-2\eta+1}(1+\bar{N}_{in}) \}}^{1/2}$ for squeezed vacuum. 
\bibitem{agarwal} G. S. Agarwal, Progress in Optics, Vol. XI, p. 33, edited by E. Wolf, North Holland, Amsterdam (1973).
\bibitem{me} S. Chaturvedi and V. Srinivasan, Journal of Modern Optics {\bf 38}, 777 (1991). }
\end{thebibliography}
\end{document}